\documentclass[epj]{svjour} \usepackage{graphics}
\usepackage{graphicx} \usepackage{epsfig} \usepackage{amsmath}
\usepackage{amssymb} \usepackage{amsfonts} \usepackage{color}
\usepackage[latin1]{inputenc} \usepackage[english]{babel}

\begin{document}

\title{Adoption of innovations with contrarians and repentant agents}

\author{Mirta B. Gordon\inst{1} \and M. F. Laguna\inst{2} \and
  S. Gon\c{c}alves\inst{3} \and J. R. Iglesias\inst{3,4}}
\institute{Univ. Grenoble Alpes, CNRS, Laboratoire d'Informatique de
  Grenoble (LIG), Grenoble, France \and CONICET - Centro At\'omico
  Bariloche, 8400 S. C. de Bariloche, R\'io Negro, Argentina.  \and
  Instituto de F\'isica, UFRGS, Caixa Postal 15051, 91501-970 Porto
  Alegre, RS, Brazil \and Programa de Mestrado em Economia, UNISINOS,
  S\~ao Leopoldo, RS, Brazil and Instituto Nacional de Ci\^encia e
  Tecnologia de Sistemas Complexos, Rio de Janeiro, RJ, Brazil.}
\date{Received:date / Revised version:date}

\abstract{The dynamics of adoption of innovations is an important
  subject in many fields and areas, like technological development,
  industrial processes, social behavior, fashion or marketing. The
  number of adopters of a new technology generally increases following
  a kind of logistic function.  However, empirical data provide
  evidences that this behavior may be more complex, as many factors
  influence the decision to adopt an innovation.  On the one hand,
  although some individuals are inclined to adopt an innovation if
  many people do the same, there are others who act in the opposite
  direction, trying to differentiate from the ``herd''. People who
  prefer to behave like the others are called mimetic, whereas
  individuals who resist adopting new products, the stronger the
  greater the number of adopters, are named contrarians.  On the other
  hand, new adopters may have second thoughts and change their
  decisions accordingly. Agents who regret and abandon their decision
  will be denominated repentant.  In this paper we investigate a
  simple model for the adoption of an innovation for a society
  composed by mimetic and contrarian individuals whose decisions
  depend mainly on three elements: the appeal of the novelty, the
  inertia or resistance to adopt it, and the social interactions with
  other agents. In the process, agents can repent and turn back to the
  old technology.  We present analytic calculations and numerical
  simulations to determine the conditions for the establishment of the
  new technology. The inclusion of repentant agents modify the balance
  between the global incentive to adopt and the number of contrarians
  who prevent full adoption, generating a rich landscape of temporal
  evolution that includes cycles of adoption.}


\maketitle

\section{Introduction}

Innovation is at the core of the changing in living conditions all
along the human history. It is also one of the main driving forces of
sustainable economical development in modern societies.  However, even
when innovations may represent a clear improvement over existing
technologies, its adoption is not guaranteed because it depends on
other factors that can restrain the adoption process, like the
individual resistance or a high price.  Besides, the adoption may be
boosted by means of advertising or interpersonal influence.
Rogers~\cite{Rogers95} was the first one to address the problem of
innovation adoption. In his qualitatively description, he claims that
adoption curves are $S$-shaped (logistic) as a function of time: there
are few early adopters, and only when their number becomes larger than
a threshold, adoption develops up to a saturation point.

Systems of heterogeneous interacting individuals are complex systems
whose properties have been studied in the contexts of economics (see
\cite{GordonNadalPhanVannimenus05,NadalPhanGordonVannimenus05,GordonNadalPhanSemeshenko14}
and references therein),
criminality~\cite{GordonIglesiasSemeshenkoNadal09}, game
theory~\cite{HauertSzabo05}, and in many other social and biological
systems.  It has been shown that when the individuals are mimetic,
i.e. they choose to imitate the behavior of the others (also called
herding or congregator behavior), the possible equilibria have well
known properties~\cite{Bass,GordonNadalPhanSemeshenko09}.  If some
individuals do not exhibit a mimetic behavior, adoption dynamics is
more involved, but also more interesting.  Such individuals, called
{\it contrarians}, have been described in different contexts in the
literature. Galam \cite{Galam08} introduced contrarian agents in voter
models, in such a way that they adopt opinions that are systematically
opposite to the one of the majority of their neighbors. Other
possibilities have been proposed more recently by
Masuda~\cite{Masuda13}, who considered different models in which the
decision of each contrarian depends on its neighborhood (made of
contrarians and/or mimetics).  It is also possible to have indifferent
agents, who are not aware of the social tendencies.  As we want to
focus on the role of repentants in the dynamics of innovation adoption
we do not consider indifferent agents in the present contribution.

With or without contrarians, the time evolution of the fraction of
adopters is a Markov chain: $n(t+1)={\cal F}(n(t))$. The fixed points
attractors of the dynamics satisfy $n={\cal F}(n)$.  However, as
demonstrated by Goles et al.~\cite{GolesEtAl85}, systems with
interacting binary agents evolve toward fixed points only when the
interactions are {\em symmetric} and positive. Negative symmetric
interactions may lead either to fixed points or to cycles of length
$2$, depending on details of the dynamics and on the initial
state. These results rely on the existence of an energy function that
is a decreasing (more rigorously, non-increasing) function of time
under the system's dynamics. However, a system with contrarians does
not necessarily have an underlying energy function, because the
interactions between mimetic agents and contrarian agents are not
symmetric. Being an individual property, contrarians have negative
interactions with all other agents. Thus, interactions between
contrarians are symmetric ---both being negative--- but interactions
between a contrarian and a mimetic agent are
anti-symmetric. Consequently, the existence of fixed points is not
guaranteed.

A ``microscopic'' model of adoption dynamics has been proposed
recently~\cite{GoncalvesLagunaIglesias12}. This model considers
heterogeneous individuals in the presence of advertising. Mimetic
individuals have a positive interaction with the adopters, increasing
their pay-off function with the fraction of them. Contrarians,
instead, have the opposite behavior, with a preference to adopt that
decreases when the fraction of adopters increases. In that model,
adopters, both mimetics and contrarians, cannot change their minds, so
that the fraction of adopters is a non-decreasing function of time.
The dynamics of adoption in models with only mimetic individuals has
been studied by Bass~\cite{Bass} and also by Phan et
al.~\cite{PhanPajotNadal03} for different types of networks. In
Ref. \cite{PhanPajotNadal03} authors have shown that the fraction of
adopters increases with time through avalanches that depend on the
underlying network structure.  Moreover, the fraction of adopters at
equilibrium in the absence of contrarians has been obtained
analytically in~\cite{GoncalvesLagunaIglesias12} for a uniform
distribution of the resistance to adopt and small values of the
interaction weights. Numerical results have been obtained when
contrarians are included: in the context of innovation the most
important consequence of the inclusion of contrarians is the
non-trivial restraining effect on the adoption curves, i.e, a small
fraction of contrarians produce a large reduction on the final
fraction of adopters.  Gon\c{c}alves et
al. model~\cite{GoncalvesLagunaIglesias12} is suitable for situations
where users can not change their decisions, such as the case of
expensive technologies.  But there are other situations, as for
example the choice of an operating system, a software, or an internet
supplier, where the decision can be revised periodically. In such
cases, adopters may change their minds and abandon the innovation.

In this article we will focus on a society where individuals exhibit a
mimetic or contrarian behavior (kept fixed during the whole adoption
process), but they can repent for their decisions, going back to a
non-adopter state.  We will study this model using analytical and
numerical approaches, and considering different distributions of the
idiosyncratic resistance to adopt.  We explore the parameters space by
comparing the results of simulations with a mean field analytic
approach, analyzing the phase diagram of the system for different
proportions of mimetics and contrarians.  The paper is organized as
follows: In section I we present the model, in section II we consider
a uniform distribution of the resistance to adopt (analytically and
numerically), and in section III we analyze the case of a logistic
distribution. Conclusions are presented in section IV.

\section{The model of adoption with social interactions}
\label{model}

We consider a system of $N$ individuals that must decide whether to
adopt or not a novelty. We define a parameter $A \ge 0$ as a global
incentive to adopt, the same for all agents. This incentive is
proportional to the advantages introduced by the new technology, to
advertising, and to eventual social values associated with the
possession of the new product. On the other hand, each individual has
a resistance to adopt the innovation given by a value $R+r_i$ ($1 \le
i \le N$), where $R$ is the population's average and the $r_i$ are
(quenched) idiosyncratic deviations distributed among the population
according to a probability density function ${\cal P}(r)$ of zero mean
and variance $s$.  In \cite{GoncalvesLagunaIglesias12} ${\cal P}(r)$
is uniform in $[-r_0,r_0]$ with $R=0.5$, $r_0=0.5$ and $s=r_0^2/3$.
This resistance can be associated to suspicion against the novelty, to
a certain laziness that induces to remain with the old technology or
to limited resources for the acquisition of the new technology.  When
confronted with the decision to adopt or not the new technology, we
assume that there are two kinds of individuals: a fraction $f$ of the
population is composed by contrarian agents, i.e., they resist to
imitate what others do, while a fraction $1-f$ of the population is
mimetic, so they tend to follow the herd. We assume these attitudes
also remain fixed during the adoption process.  At each time step,
each agent weights the {\em expected} decisions of the others with a
strength $J_i$, which represents the social influence on his own
decision.  In \cite{GoncalvesLagunaIglesias12}, mimetic individuals
{\em increase}, while contrarians {\em decrease}, their willingness to
adopt the innovation proportionally to the fraction of adopters $n$,
with weight $J_i=J=1$ for all $i$. In other models
\cite{Galam08,Masuda13}, individuals are susceptible to the {\em
  majority}, meaning that the willingness to adopt are proportional to
$n-1/2$ instead of $n$.

As it stands, the model has four parameters: $A$, $R$, $J$, and
$s$. We can get rid of $J$ by using it as a normalization factor for
the others, as in
\cite{GordonNadalPhanVannimenus05,GordonNadalPhanSemeshenko09}.  In
the following we measure all the values in terms of the strength of
the social interactions, $J$, and define
\begin{equation}
d \equiv \frac{A-R}{J} \; , \; u_i \equiv \frac{r_i}{J} \; , \; u_0
\equiv \frac{r_0}{J}\; , \, \sigma \equiv \frac{s}{J}
\end{equation}
Notice that this normalization was implicit in
\cite{GoncalvesLagunaIglesias12}, where most of the time it was
assumed that J=1.

The pay-offs of the model \cite{GoncalvesLagunaIglesias12} are:
\begin{subequations}
\label{eq:payoffsModel1}
\begin{align}
\pi_i^M = d-u_i+n_{-i} \;\;\;\; &{\rm if} \;\; i \;\; {\rm is \;\;
  mimetic,} \\ \pi_i^C = d-u_i-n_{-i} \;\;\;\; &{\rm if} \;\; i \;\;
   {\rm is \;\; contrarian,}
\end{align}
\end{subequations}
where $n_{-i}$ is the fraction of adopters without counting individual
$i$:
\begin{equation}
n_{-i}=\frac{1}{N-1} \sum_{k \ne i} \omega_k,
\label{eq:n}
\end{equation}
with $\omega_k = 1$ if $k$ is an adopter and zero otherwise.

When calculating the expected pay-offs we can check each individual at
random and immediately update his decision according to the sign of
his pay-off. In this case the expected pay-off is equal to the actual
one. On the other hand we can perform a synchronous updating of all
agents at once. In the later case agents determine their pay-offs
taking into account the last value of the number of adopters, $n$, but
this value is refreshed after all agents have been checked. So, the
expected and the actual pay-offs may be different after the
updating. We will discuss this point in further detail in the next
section.

If adopters are not allowed to change their minds, as in the model
considered in \cite{GoncalvesLagunaIglesias12}, only the actual
pay-offs of non-adopters are important for the dynamics; however, the
equilibrium properties of the present model depend both on adopters
and non-adopters pay-offs. In the limit of large populations with
large numbers of adopters we may replace $N-1 \approx N$, drop down
the constraint $k \neq i$ in the equation above and approximate
$n_{-i}$ by the bare fraction of adopters $n$:
\begin{equation}
\label{eq:ni}
n_{-i} \approx n \equiv \frac{1}{N} \sum_{k} \omega_k.
\end{equation}

Individuals adopt the new technology whenever their expected pay-offs
are positive.  The adoption dynamics may become quite complex upon
introduction of contrarians that decide according to the majority
rule~\cite{GoncalvesLagunaIglesias12}.  Here we include the
possibility of coming back from previous decisions, thus, individuals
will abandon innovation if the pay-off is negative. No doubts or
delays are allowed in this version of the model.  The results of the
present and the previous~\cite{GoncalvesLagunaIglesias12} models are
compared and discussed in the following, that is, when adoption can be
reverted or not.

When considering a large number of agents we can take the limit $N
\rightarrow \infty$. Introducing the fraction of adopters (\ref{eq:n})
in equations (\ref{eq:payoffsModel1}), and assuming that the
idiosyncratic normalized resistance to adopt $u_i$ are quenched random
variables of probability density ${\cal P}(u)$, the adoption
probability (the probability of positive pay-off) in the limit $N
\rightarrow \infty$ is
\begin{subequations}
\label{eq:adoptionProba}
\begin{align}
P(\omega=1|M) =&\int_{-\infty}^{d+ n} {\cal P}(u) du \\ P(\omega=1|C)
=& \int_{-\infty}^{d-n} {\cal P}(u) du
\end{align}
\end{subequations}
where $M$ stands for mimetic and $C$ for contrarian agents.

If the fraction of adopters at time $t$ is $n(t)$, the adopters'
dynamics is given by the following equation:
\begin{eqnarray}
\label{eq:dynamics}
n(t+1) &=& (1-f)\int_{-\infty}^{d+ n(t)} {\cal P}(u) du + f
\int_{-\infty}^{d-n(t)} {\cal P}(u) du \nonumber \\ &=&
\int_{-\infty}^{d+ n(t)} {\cal P}(u) du -f \int_{d-n(t)}^{d+n(t)}
    {\cal P}(u) du
\end{eqnarray}

In the absence of contrarians, $f=0$, the phase diagram of the model
is well known
\cite{GordonNadalPhanVannimenus05,NadalPhanGordonVannimenus05,GordonNadalPhanSemeshenko09}. The
stationary states satisfy
\begin{equation}
\label{eq:stationaryNoContrarian}
n = \int_{-\infty}^{d+ n} {\cal P}(u) du
\end{equation}
which is easily solved (see \cite{GordonNadalPhanSemeshenko09}).

In the following sections we include contrarians and repentants, we
assume pay-offs given by equations (\ref{eq:payoffsModel1}) and we
include the possibility of changing decisions for both mimetic and
contrarian agents. Agents adopt if the expected pay-off is positive
and do not adopt otherwise. Moreover, those that have adopted
previously may go back to no-adoption (repentants) if the actual
pay-off turns out to be negative.  We analyze two particular
distributions ${\cal P}(u)$, the uniform distribution and the logistic
one.

\section{Uniform distribution}
\label{sec:UD}
In this section, we present first results of simulations and then we
discuss analytic results for a uniform distribution, i.e. ${\cal
  P}(u)=(2 \, u_0)^{-1}$ in $[-u_0,u_0]$ and ${\cal P}(u)=0$
elsewhere. We will compare the results of this model with the ones
presented in \cite{GoncalvesLagunaIglesias12}, thus we adopt the same
value of the parameters of that paper: i.e. we consider $u_0=0.5$ so
that ${\cal P}(u)=1$. We also restrict the comparison to the results
in ref. \cite{GoncalvesLagunaIglesias12} where $J=1$, so we can
consider the variables already normalized.

\subsection{Numerical results}
When performing the simulations we consider two different dynamics,
corresponding to synchronous and non-synchronous updates. In the case
of synchronous Parallel Dynamics (PD), the pay-offs are evaluated for
all agents at the same time, then the status of each agent is changed
or not accordingly to their pay-off and thereafter the new fraction of
adopters, $n$, is updated. In the Monte Carlo sequential dynamics
(MC), one agent is selected at random and its status is updated
depending on its pay-off, then the number of adopters is immediately
adjusted. This process is repeated $N$ times, which corresponds to one
MC step.  The reason for considering these two dynamics is that the
first one (PD) is better adapted to be compared with analytical
results, while MC simulations probably provides a better description
of the changes in real societies. The difference between the two
dynamics is that in the MC method a sequential update is performed,
which means that the number of adopters changes in a continuous way
during each MC step, while in the PD case the number of adopters is
updated at the end of each step, after evaluating the
pay-offs of all agents. A second difference between the two types of
dynamics is that with the PD procedure all agents are visited, while
in MC dynamics they may not.

\begin{figure}[htbp]
	\centering \includegraphics[width=0.45\textwidth]{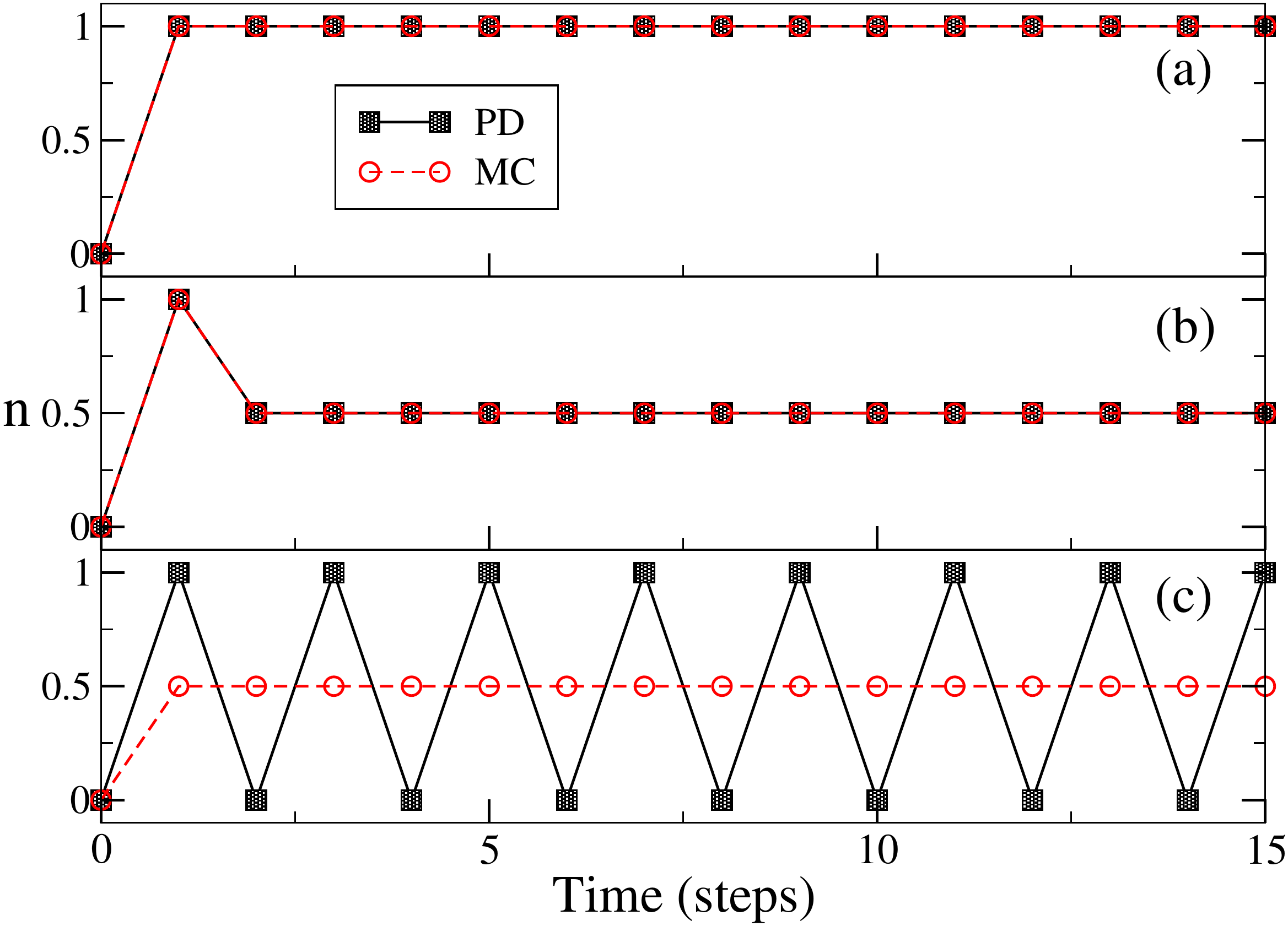}
	\caption{Case of two agents: (a) Two mimetics, the final value
          of $n$ is $n=1$; (b) one mimetic and one contrarian, $n$
          quickly converges to $n=0.5$; (c) two contrarians, the
          system exhibits oscillations.  Black squares correspond to
          PD and open red circles to MC dynamics. In all cases,
          $d=0.01$ and $u_1=u_2=0$.}  \qquad\qquad
	\label{fig:twoagents}
\end{figure}

In order to illustrate the dynamics let us first consider a very
simple case with just two agents, $N=2$, and the three possible
combinations: two mimetics, one mimetic and one contrarian, and two
contrarians. We choose parameters such that for both agents $d$ is
slightly higher than $u_i$, for example $d=0.01$ and $u_1=u_2=0$. The
results are exhibited on Fig.~\ref{fig:twoagents} (black filled
squares correspond to PD and red open circles to MC dynamics).  In the
case of two mimetic agents, both of them will adopt immediately and no
further changes are observed, the system arrives at a fixed
point. With one mimetic and one contrarian, both agents adopt in the
first time step but then only the mimetic remains as adopter. The
system gets to a fixed point with $n=0.5$. In the cases above the
description corresponds strictly to PD while some random variations
are possible with MC dynamics.  An interesting time evolution arises
when there are two contrarians because both of them adopt when $n=0$,
but as soon as they adopt, $n_{-i}=1/2$ (notice that for small values
of $N$, the approximation given by Eq.~\ref{eq:ni} is not valid), so the
contrarian's pay-off becomes negative and at the next evaluation both
become non-adopters.  Therefore, the system exhibits a strictly
periodic behavior in parallel simulations. In MC simulations, however,
there are no oscillations: after the first agent adopts, when the
second is selected, it will not adopt because its pay-off will be
negative, so the evolution stops at 50\% of adopters.

\begin{figure}[htbp]
	\centering \includegraphics[width=0.45\textwidth]{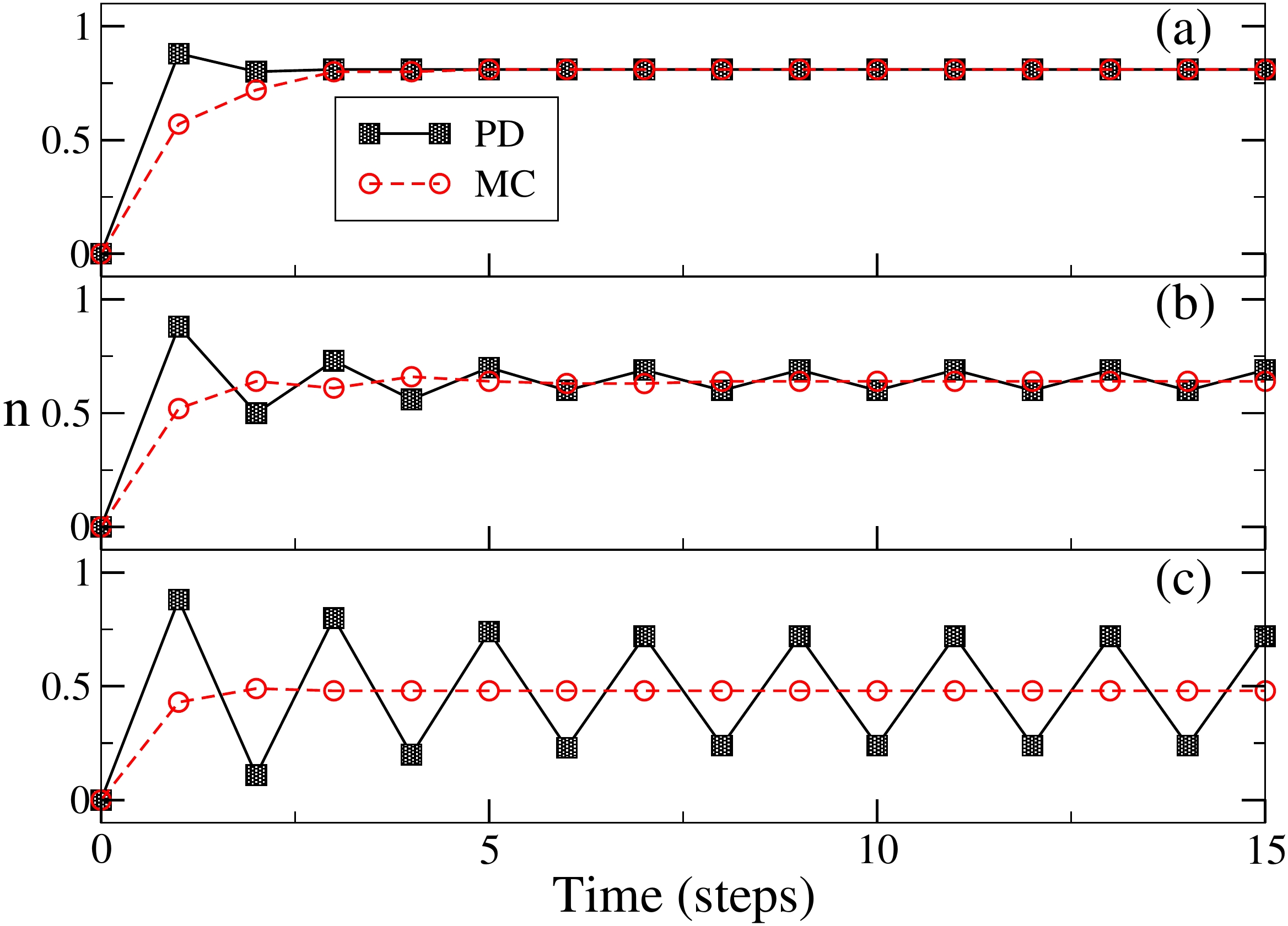}
    	\caption{Fraction of adopters as a function of time for
          $N=100$ agents and different values of the fraction of
          contrarians $f$: (a) $f=0.2$, no oscillations; (b) $f=0.5$,
          small amplitude oscillations; and (c) $f=0.9$, large
          amplitude sustained oscillations.  Black squares correspond
          to PD and open red circles to MC dynamics. In all cases,
          $d=0.4$ and $u_0 = 0.5$.} \qquad \qquad
\label{100agents}	
\end{figure}

 Let us now consider the case with an intermediate number of agents,
 $N=100$; henceforth we use the approximation given by Eq.(4). In
 Fig.~\ref{100agents}, it can be verified that when the fraction of
 contrarians is large, $f=0.9$, the system exhibits oscillations in
 the parallel dynamics case, while there are no oscillations with
 Monte Carlo dynamics. For a lower fraction of contrarians, $f=0.5$,
 oscillations are of smaller amplitude and disappear for $f \leq 0.2$.

\begin{figure}[htbp]
	\centering \includegraphics[width=0.45\textwidth]{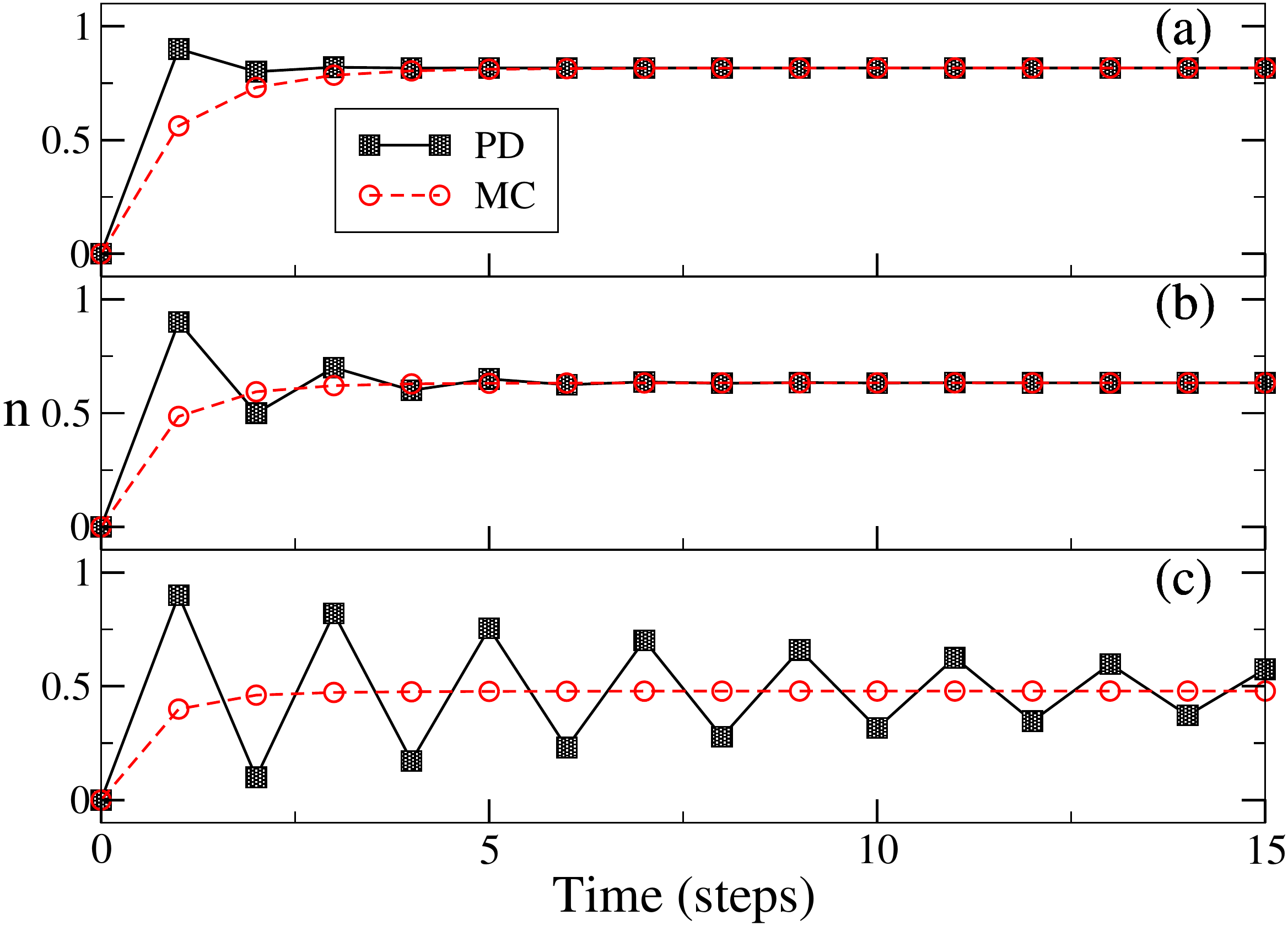}
	\caption{Fraction of adopters as a function of time for
          $N=10^{7}$ agents and different values of the fraction of
          contrarians $f$: (a) $f=0.2$, no oscillations; (b) $f=0.5$,
          very short lived oscillations; and (c) $f=0.9$, transient
          oscillations.  Black squares correspond to PD and open red
          circles to MC dynamics. The other parameters in all cases
          are $d=0.4$ and $u_0 = 0.5$.} \qquad \qquad
	\label{muchos}
\end{figure}

For an even larger number of agents, $N=10^{7}$, oscillations always
disappear in the long term. Even for a high proportion of contrarians,
$f = 0.9$, and with parallel dynamics, oscillations decay after a
short transient, as can be seen in Fig.~\ref{muchos}.  For a lower
fraction of contrarians, oscillations are very short lived; for
$f=0.5$, for instance, no more than three oscillations are seen in
Fig.~\ref{muchos}b.  The previous results, all put together, suggest
that for uniform distribution of resistance to adoption ---which is a
raw simplification of the society representation---, oscillations are
possible for a large number of contrarians, but are a finite size,
transient effect. We have investigated the existence, or not, of
oscillations, and it seems that the critical number of agents is
around $N \approx 10000$.

\begin{figure}[htbp]
	\centering \includegraphics[width=0.45\textwidth]{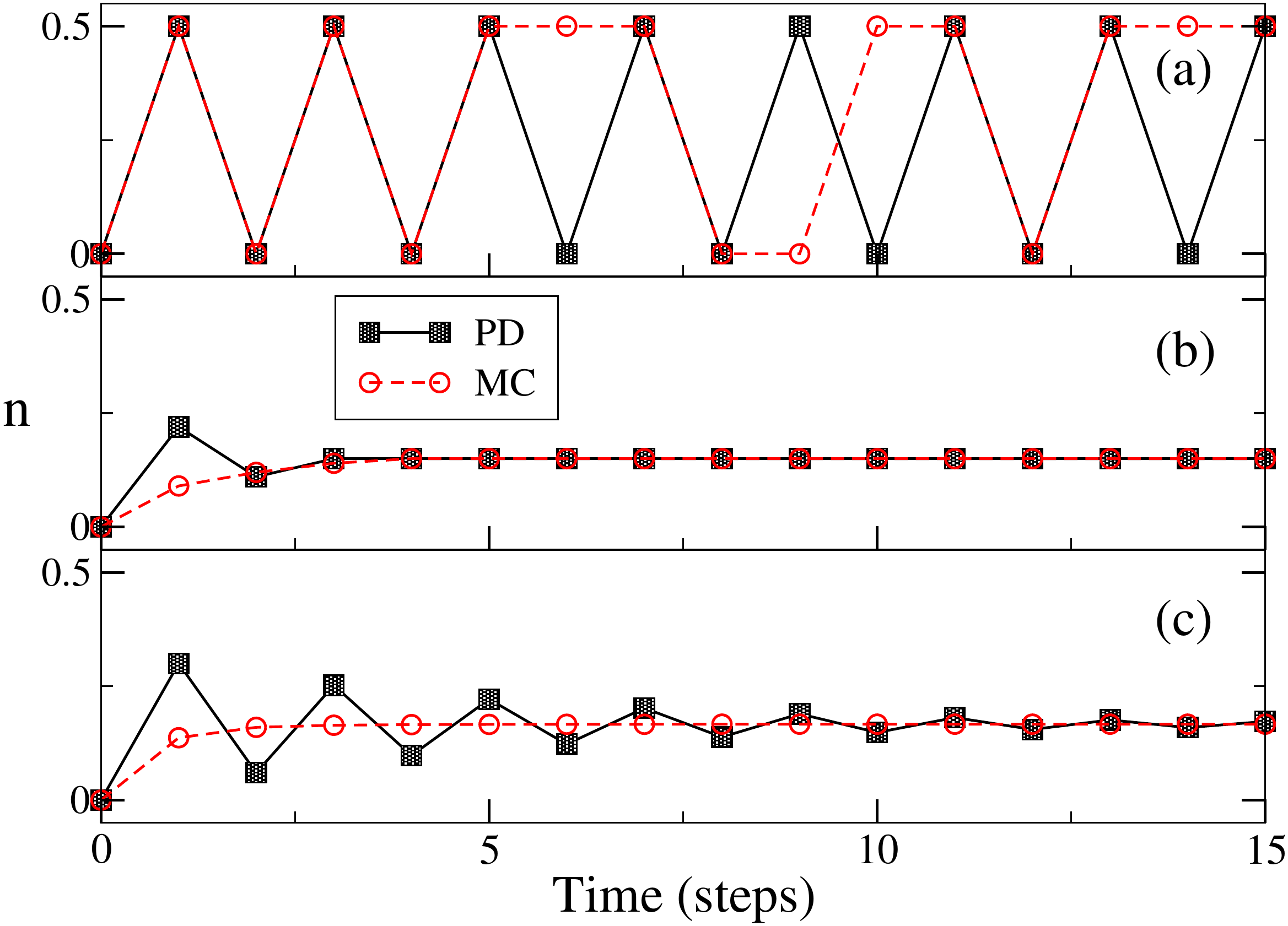}
	\caption{Fraction of adopters as a function of time for a
          large fraction of contrarians ($f=0.9$), but a low value of
          advertising ($d=-0.2$). Results for different system sizes:
          (a) $N=2$, (b) $N=100$, and (c) $N=10^{7}$. Black squares correspond to
          PD and open red circles to MC dynamics. The qualitative
          behavior is the same as in Fig.~\ref{100agents} and
          \ref{muchos} with the same value of $f$, but a bigger value
          of $d$.} \qquad \qquad
        \label{lowd}
\end{figure}

It is also interesting to investigate the effect of the
advertising. The values considered, $d=0.4$ may be too high. It can be
argued that a too high value of the incentive to adopt could have a
role in the appearance, or not, of oscillations. And it has. In order
to check this we try a lower value of the incentive, $d=-0.2$. As $u$
goes from $-0.5$ to $+0.5$ that value of $d$ implies that $30\%$ of
the agents have an idiosyncrasy below $d$, i.e. $30\%$ are potential
early adopters.  As we are interested in possible oscillations we
focus on a high concentration of contrarians, $f=0.9$, and three
system sizes, $N=2$, $N=100$, and $N=10^{7}$.  The results are shown
in Fig.~\ref{lowd} where a clear feature can be seen: the asymptotic
values for the average fraction of adopters are much lower ($n \approx
0.15$) than in the case of $d=0.4$. This is expected because the
advertising is what promotes the adoption in the first place. But on
the other hand the oscillatory behavior is very similar to previous
results with a bigger value of $d$, so our conclusions regarding the
oscillatory behavior (and the existence or not of oscillations) are
robust against a change of the advertising. However, if the width of
the distribution is narrower, no oscillations appear for low values of
$d$.

Note that the effect of the ``repentants" with the associated cycle
dynamics is only possible if $f > 0$. If there are no contrarians,
even if it is possible to repent, no agent will do it because for a
mimetic the pay-off can not decrease.  In other words, cycles are only
possible if both, contrarians and repentants, are present in the
system. Moreover, the effect of contrarians modifies the number of
adopters, and this change may induce mimetics to abandon the
innovation.  We finally remark that in this section we have restrained
our simulations to the parameters utilized in
ref.~\cite{GoncalvesLagunaIglesias12}. In particular, we have chosen
$u_0=0.5$. If a narrower distribution of the idiosyncratic resistance
to adopt is considered, stable oscillations may appear for relatively
high values of the advertising. In the next section we show, as an
example, that such is the case for $u_0 = 0.25$ and ${\cal P}(u)=2$
(see Fig.\ref{fig:extremes-b}). Moreover, the effect of the width of
the distribution on the existence of oscillations will be discussed in
detail in section \ref{logistic}, as the logistic distribution is
simpler to be treated.

\subsection{Analytic Results}
\label{analytical}

In this section we present analytic mean field results and compare
them with numerical simulations of the preceding section.  As the
analytic calculations implicitly assume $N \rightarrow \infty$ we will
compare them with the numerical results for the big size system, i.e.,
$N=10^{7}$.

\begin{figure}[htbp]
	\centering \includegraphics[width=0.45\textwidth]{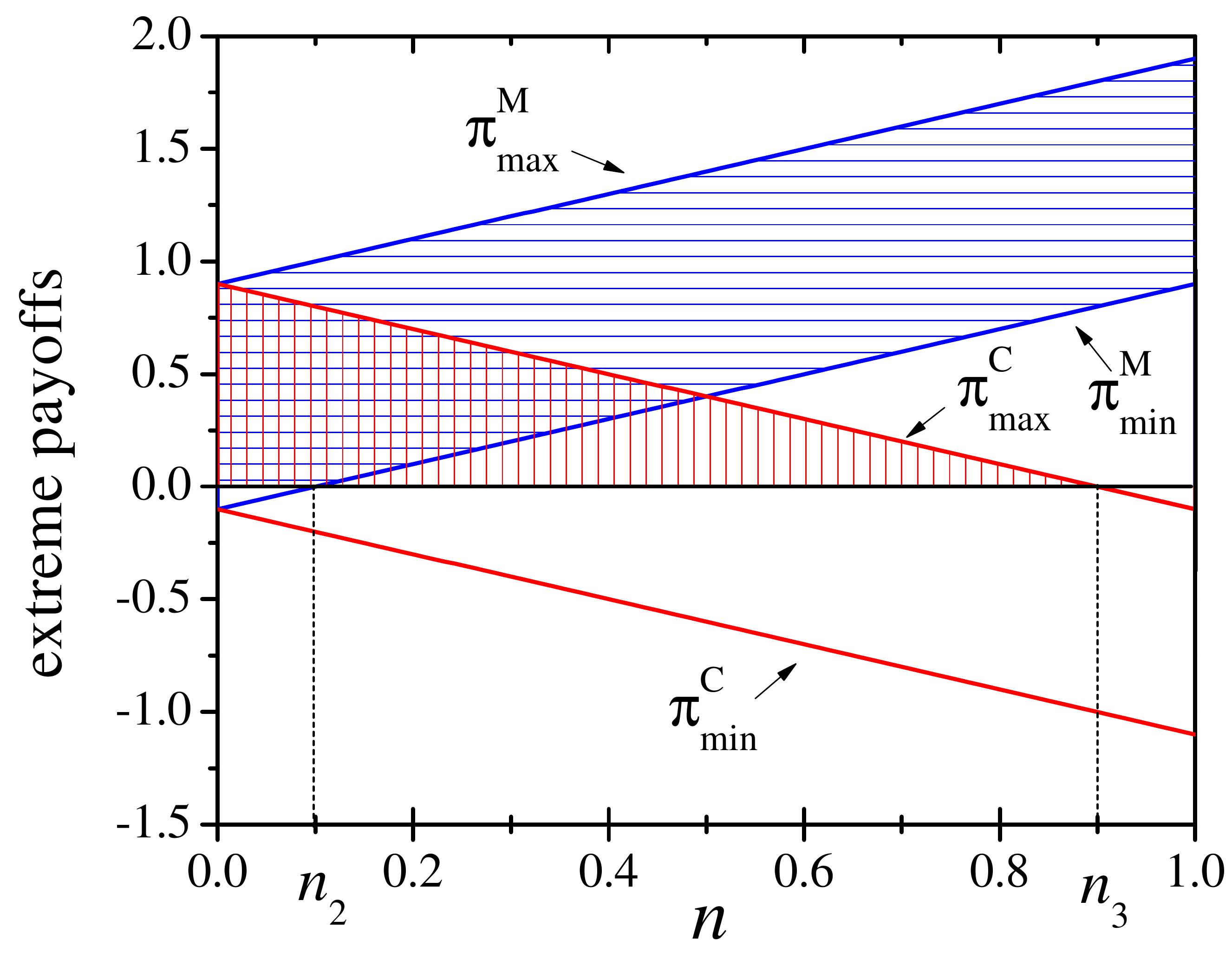}
	\caption{Extreme pay-offs as a function of the number of
          adoppters,$n$, for the uniform distribution with $d=0.4$,
          $u_0 = 0.5$. $\pi$ lines are extreme pay-offs, blue lines
          for mimetics and red lines for contrarians. In general they
          are given by Eq.~\ref{eq:MaxMinPayoffs}, in this case by
          Eq.~\ref{eq:MaxMinPayoffs2}.} \qquad \qquad
	\label{fig:extremes}
\end{figure}

Due to the compact support of ${\cal P}$, the possible (normalized)
pay-offs as a function of $n$ (see Eq.~\ref{eq:payoffsModel1} and
Fig.~\ref{fig:extremes}) are bounded by
\begin{equation}
  \begin{aligned}
    \label{eq:MaxMinPayoffs}
    \pi_{max}^M(n) &= d+u_0+ n \\
    \pi_{min}^M(n) &= d-u_0+ n \\
    \pi_{max}^C(n) &= d+u_0- n \\
    \pi_{min}^C(n) &= d-u_0- n
  \end{aligned}
\end{equation}
and the fixed point equations of the dynamics, $n(t+1)=n(t)=n$ are:
\begin{subequations}
\label{eq:fixed point}
\begin{align}
n =& n^M+n^C; \label{eq:fixed point_n} \\ n^M =& (1-f)
\int_{max[0,\pi_{min}^M(n)]}^{max[0,\pi_{max}^M(n)]} {\cal P}(u)
du \label{eq:fixed point_M} \\ n^C =& f
\int_{max[0,\pi_{min}^C(n)]}^{max[0,\pi_{max}^C(n)]} {\cal P}(u)
du \label{eq:fixed point_C}
\end{align}
\end{subequations}
that must be solved for $n$. We call $n^M$ the number of adopters who
are mimetic and $n^C$ those who are contrarian.

There are different regimes that have to be analyzed separately,
depending on the signs of the extreme pay-offs at $n=0$ and at
$n=1$. To illustrate this point we show on Fig. \ref{fig:extremes} the
extreme pay-offs as a function of the number of adopters, $n$, for
$d=0.4$ and $u_0=0.5$. Blue lines are extreme pay-offs for mimetics
and red lines for contrarians. The area with positive pay-off
correspond to the number of mimetics (blue ones) and contrarians (red
ones) but one should make attention to the fact that these areas are
weighted with the factors $(1-f)$ for mimetics, and $f$ for
contrarians. When $f=0.5$ both areas have the same weight. In this
case, starting with $n=0$ the number of adopters after the first step
is of the order of $n=0.9$. But with such a high number of adopters
most of the contrarians will defect the innovation and the number of
adopters will fall to $n=0.5$ in the second step. After that, the
number of adopters increases again, and then decreases to finally
converge to an intermediary value of $n=0.6$. These decaying
oscillations converging to $n=0.6$ are also observed in the numerical
results, on Fig. \ref{muchos}(b)

We consider now the points where the extreme pay-offs change sign by
solving Eqs.~\ref{eq:fixed point} for the pay-off equal to zero:

\begin{equation}
  \begin{aligned}
    \label{eq:intersections}
    n_1 &=& -d-u_0 \\
    n_2 &=& -d+u_0 \\
    n_3 &=& d+u_0 = -n_1 \\
    n_4 &=& d-u_0 = -n_2
  \end{aligned}
\end{equation}

Two of these points, $n_2$ and $n_3$, are also indicated on
Fig. \ref{fig:extremes}; $\pi_{max}^M(n)$ is always positive and
$\pi_{min}^C(n)$ always negative, so $n_1$ and $n_4$ are both negative
and are not solutions.  In the general case we can state that $u_0 >
0$, $n_2 > n_1$ and $n_3 > n_4$, but depending on the relative values
of $d$ and $u_0$, $n_3$ may be larger or smaller than $n_2$. Here we
would like just to analyze the two cases that we have simulated
numerically: $d=0.4$ and $d=-0.2$, both with $u_0 = 0.5$. In the first
case ($d=0.4$) one has the following boundaries:

\begin{equation}
  \begin{aligned}
    \label{eq:MaxMinPayoffs2}
    \pi_{max}^M(n) &=& 0.9+ n \\
    \pi_{min}^M(n) &=& -0.1+ n \\
    \pi_{max}^C(n) &=& 0.9- n  \\
    \pi_{min}^C(n) &=& -0.1- n
  \end{aligned}
\end{equation}

Those boundaries are the ones plotted on Fig.~\ref{fig:extremes}. The
boundary $\pi_{max}^M(n)$ is always positive, $\pi_{min}^M(n)$ is
positive for $n>0.1$, $\pi_{max}^C(n)$ is positive if $n<0.9$ and
$\pi_{min}^C(n)$ is always negative, so in the corresponding
contrarian integral (eq.~\ref{eq:fixed point_C}) the lower bound is
always $0$. The fixed point can be evaluated using Eqs.~\ref{eq:fixed
  point}. Considering $f=0.5$ in those equations, it is easy to verify
that the equilibrium is in the region $0.1 \leq n \leq 0.9$ and the
result is $n\approx 0.63$, that coincides very well with the
asymptotic limit showed in Fig.~\ref{muchos}b. For $f=0.9$ the
asymptotic value is $n \approx 0.48$ that also coincides with the
numerical result (Fig.~\ref{muchos}c). The figure also explains the
oscillations before attaining the fixed point, as described above.

In the case with $d=-0.2$, the boundaries are:

\begin{equation}
  \label{eq:MaxMinPayoffs3}
  \begin{aligned}
    \pi_{max}^M(n) &=& 0.3+ n  \\
    \pi_{min}^M(n) &=& -0.7+ n \\
    \pi_{max}^C(n) &=& 0.3- n  \\
    \pi_{min}^C(n) &=& -0.7- n
  \end{aligned}
\end{equation}

Now, the boundary $\pi_{max}^M(n)$ is always positive,
$\pi_{min}^M(n)$ is positive for $n>0.7$, $\pi_{max}^C(n)$ is positive
if $n<0.3$ and $\pi_{min}^C(n)$ is always negative. Let's examine the
case $f=0.9$. If one assumes a trial value, $n_t$, restricted to $n_t
> 0.3$ there are no contrarians adopting and the number of adopters
should be $n= (1-f) n_t$ that is always lower than $n_t$, in
contradiction with the hypothesis. So, $n$ must be lower than $0.3$
and the solution is $n \approx 0.16$ again in agreement with the
simulations (See Fig.~\ref{lowd}c).

\begin{figure}[htbp]
	\centering
        \includegraphics[width=0.45\textwidth]{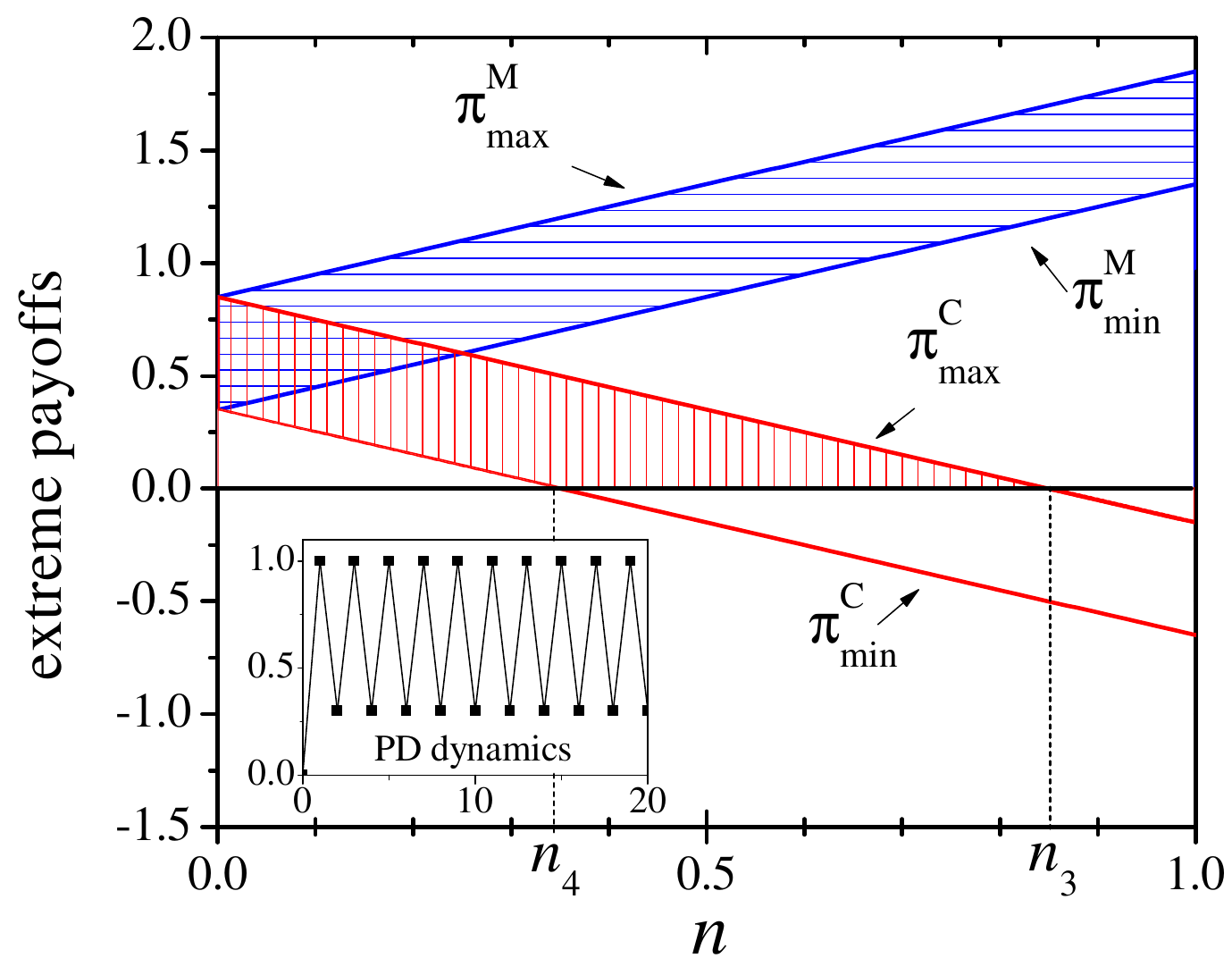}
	\caption{Extreme pay-offs as a function of the number of
          adopters, $n$, for the uniform distribution with $d=0.6$,
          $u_0 = 0.25$. $\pi$ lines are extreme pay-offs, blue lines
          for mimetics and red lines for contrarians. In general they
          are given by Eq.~\ref{eq:MaxMinPayoffs}. Inset: Numerical
          results of the fraction of adopters as a function of time in
          the parallel update dynamics for $N=10^{7}$ ($d=0.6$, $u_0 =
          0.25$, and $f=0.7$).} \qquad \qquad
	\label{fig:extremes-b}
\end{figure}

Finally, and as we said, sustained oscillations can be obtained for a
set of parameters such that $0<n_4\equiv d-u_0<n_3\equiv d+u_0<1$,
provided that both, $u_0$ and $d$ be small enough. We show in
Fig.~\ref{fig:extremes-b} the extreme pay-offs obtained analytically
for $d=0.6$ and $u_0 = 0.25$. The interpretation of this figure is the
same as the one done for Fig.~\ref{fig:extremes}. Moreover, the inset
of Fig.~\ref{fig:extremes-b} shows the numerical result for the same
values of parameters and $f=0.7$. It can be observed that the fraction
of adopters as a function of time for parallel update dynamics present
oscillations, in accordance with the analytical result.

\section{Logistic distribution}
\label{logistic}

While the uniform distribution is simpler than other distributions,
the discontinuity at the borders generates some complications
particularly for the analytic calculations. Also, one may imagine that
the distribution of idiosyncrasies in a real society exhibit a
concentration of values around the mean value and a relatively low
concentration in the extremes. Taking these points into consideration
one could envisage the use of a Gaussian distribution, as most of the
values of the resistance to adopt will be distributed within a
bell-shape of a few standard deviations width. However, the integral
of the Gaussian is not analytic. So, to avoid the cumbersome
complications raised by the uniform and Gaussian distributions, we
consider hereafter a bell-shaped logistic distribution of the
resistances to adopt, ${\cal P}(u_i)$, that is continuous and has
infinite support. The probability density of the logistic distribution
is:

\begin{equation}
{\cal P}(u) = \frac{\beta}{2 \, {\rm cosh}^2(\beta\,u)}.
\label{eq:logis}
\end{equation}

\noindent with its variance given by

$\sigma = \frac{\pi}{2 \, \beta \, \sqrt{3}}.$

Following the practice of the previous section, we present first the
numerical results.

\subsection{Numerical Results}

\begin{figure}[htbp]
	\centering
        \includegraphics[width=0.45\textwidth]{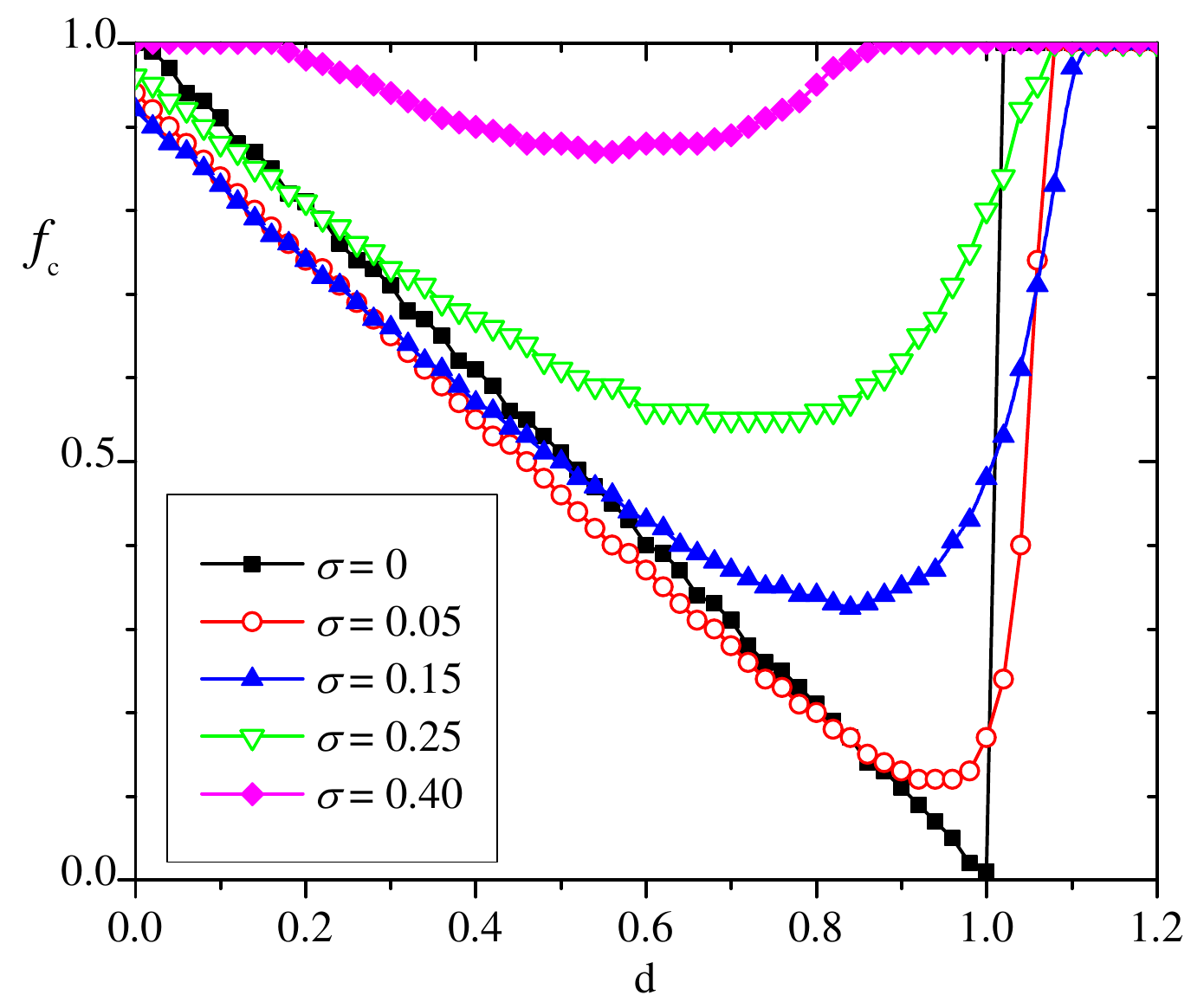}
	\caption{Thresholds value of the fraction of contrarians above
          which oscillations appear. The curves correspond to
          different values of the width of the logistic distribution
          $\sigma$, as indicated. We have represented just positive
          values of $d$ as there are no oscillations for negative
          values.  The curves go through a minimum that is lower the
          narrower the distribution. Notice that the oscillations
          disappear when the advertising is slightly higher than
          $d=1$} \qquad \qquad
	\label{umbral}
\end{figure}

\begin{figure}[htbp]
	\centering \includegraphics[width=0.45\textwidth]{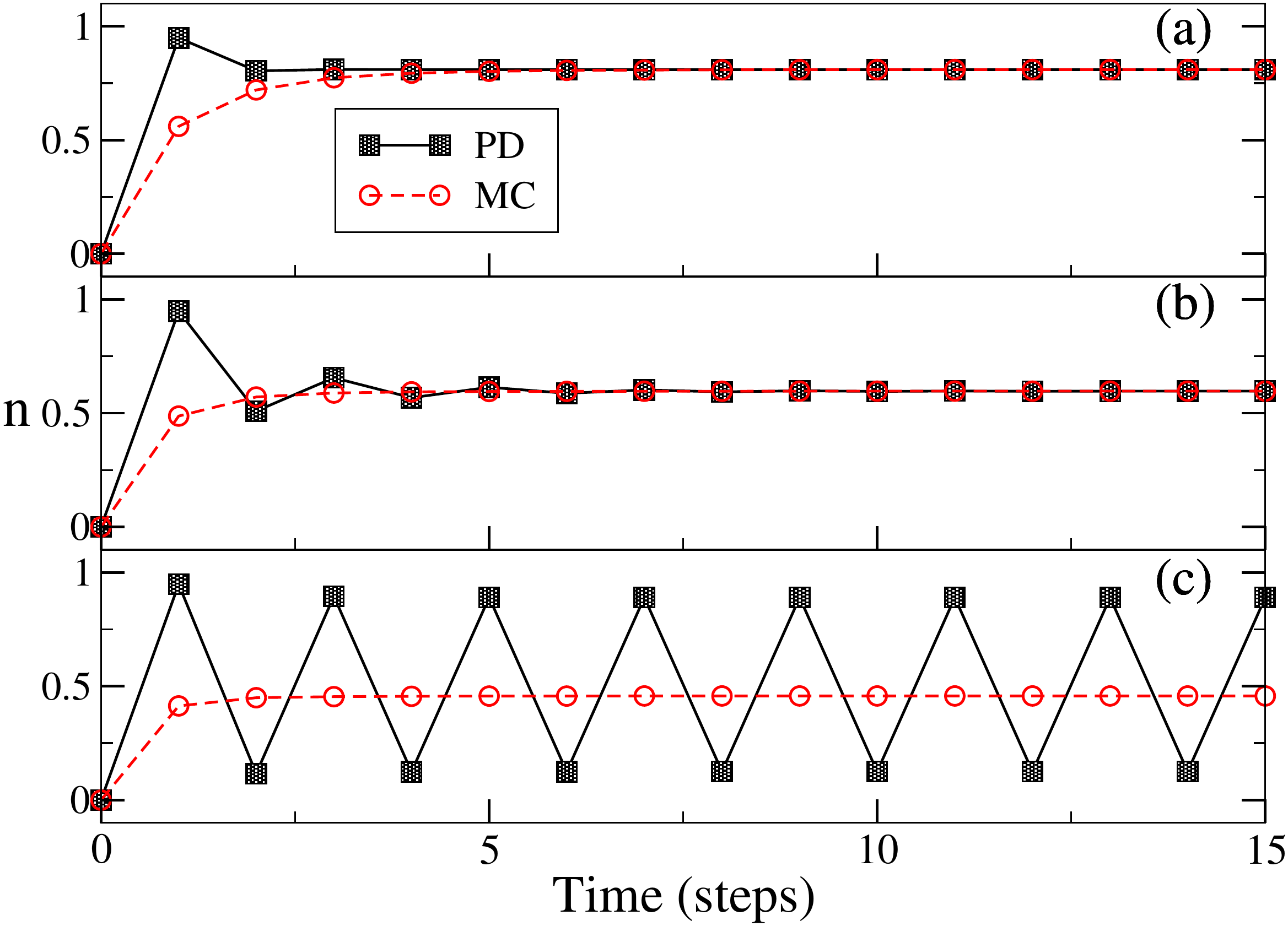}
	\caption{Temporal behavior of the fraction of adopters for the
          logistic distribution with $\sigma = 0.25$, $d=0.4$, and
          different values of the fraction of contrarians $f$: (a)
          $f=0.2$, (b) $f=0.5$, and (c) $f=0.9$. Results are for a
          large number of agents, $N=10^{7}$.  Open red circles
          correspond to the MC simulations and black squares to PD
          simulations. In the PD case it is possible to see the
          oscillations in the number of adopters for a high
          concentration of contrarians. We have considered much longer
          times than those represented in the figure and the
          oscillations are stable.}  \qquad \qquad
	\label{logd04}
\end{figure}

\begin{figure}[htbp]
	\centering \includegraphics[width=0.45\textwidth]{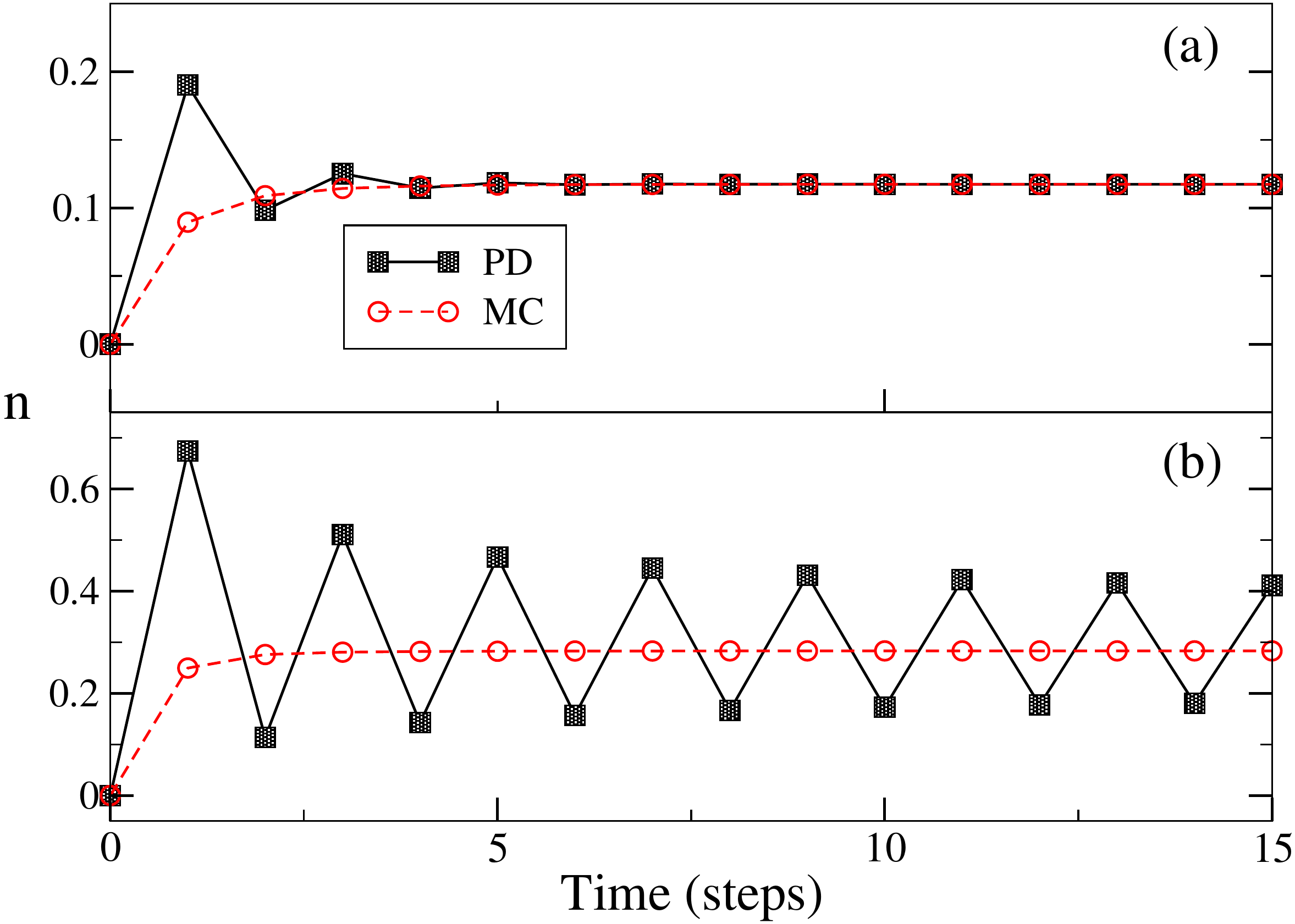}
	\caption{Temporal behavior for the logistic distribution with
          $\sigma = 0.25$, $f=0.9$ and two different values of the
          parameter $d$ (the normalized effective marketing): (a)
          $d=-0.2$ and (b) $d=0.1$. Oscillations in the number of
          adopters are obtained if $d > 0$.}  \qquad \qquad
	\label{logd03}
\end{figure}

\begin{figure}[htbp]
	\centering
        \includegraphics[width=0.45\textwidth]{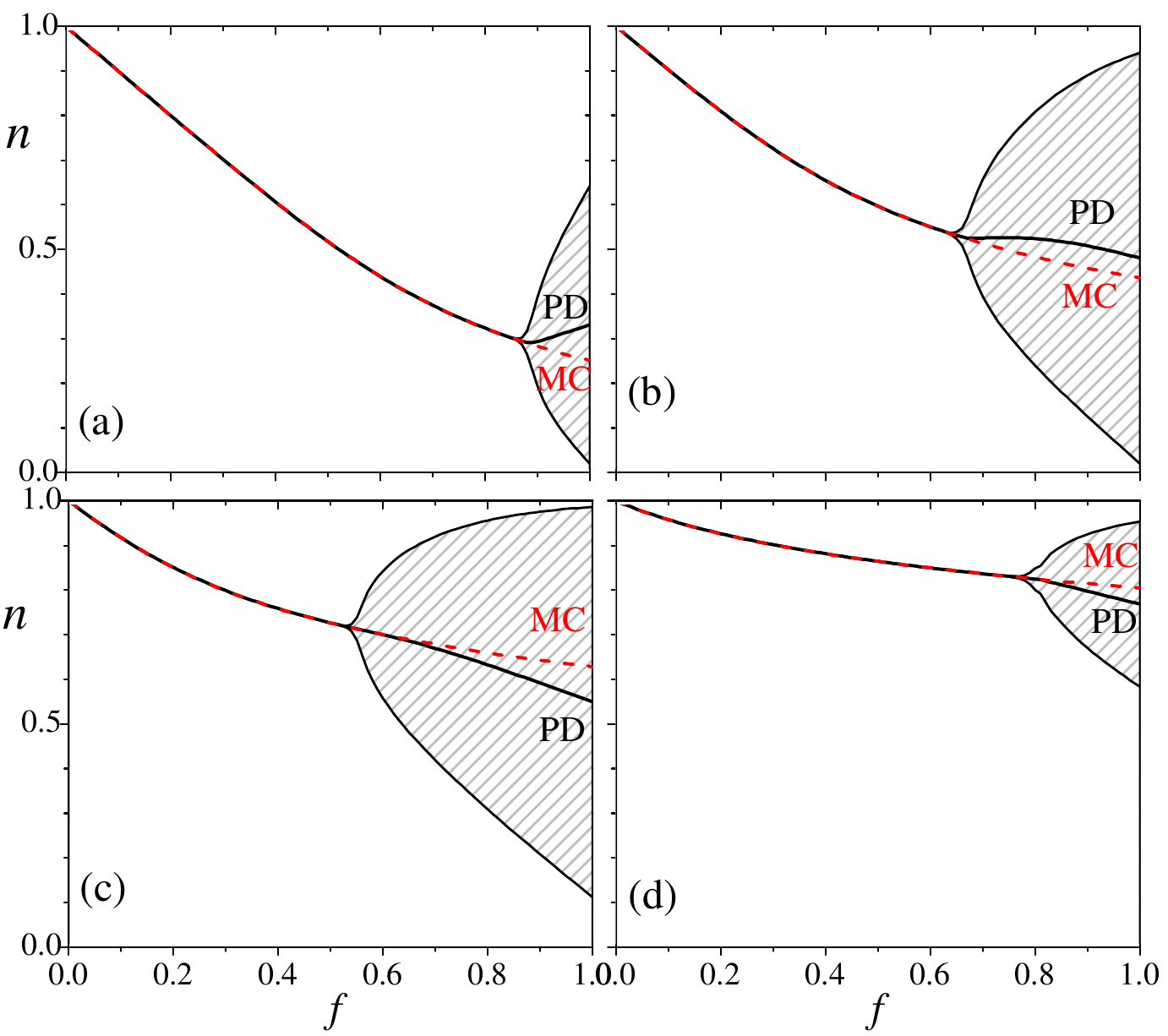}
	\caption{This figure summarizes the numerical results for the
          logistic distribution of idiosyncrasies with $\sigma =
          0.25$. All four panels exhibit the fraction of adopters as a
          function of the fraction of contrarians for four different
          values of $d$: (a) $d=0.1$, (b) $d=0.4$, (c) $d=0.7$, and
          (d) $d=1.0$. As expected, the number of adopters decreases
          when the number of contrarians increases. The red curves
          (dashed) correspond to Monte Carlo simulations and the black
          ones to a parallel dynamics. An oscillatory behavior is
          obtained only for parallel dynamics and the black lines
          correspond to the average value of the oscillations, while
          the shadowed areas indicates the amplitude of the
          oscillations. Both dynamics exhibit identical results for
          low and intermediate values of $f$, but there exists a
          critical value of $f$ when the parallel dynamics exhibits
          period two oscillations. When increasing $d$ the region of
          oscillations increases up to $d=0.7$ and then decreases for
          $d=1.0$. When $d < 0$ there are no oscillations and both
          dynamics produce the same results.}  \qquad \qquad
	\label{logres}
\end{figure}

We have performed different simulations with the logistic distribution
given by Eq.~\ref{eq:logis}. The results are presented in
Figs.~\ref{umbral}, \ref{logd04}, \ref{logd03}, and \ref{logres}. When
the simulation is performed in parallel (PD), permanent oscillations
may appear. This is the case for intermediate values of the
advertising, $d$, and relatively high values of the fraction of
contrarians, $f$. This can be verified in Fig.~\ref{umbral} where we
have represented a threshold value of the fraction of contrarians
$f_c$, above which oscillations appear, as a function of the
advertising $d$. Notice that there are no oscillations for $d<0$ or
for high values of $d$. Oscillations are present in a region of values
of $d$ around $d=1$, i.e. when the advertising is as strong as the
social interaction; oscillations are cycles of period two and arises
because contrarians adopt when the number of adopters is low, but
abandon the innovation when the number of adopters is high. However
some mimetics may follow the contrarian's behavior. The amplitude of
the oscillations decreases when decreasing $f$ or when $d$ is smaller
or bigger than $1$ (See Fig. \ref{umbral}). The region where stable
oscillations occur is larger the narrower the width of the
distribution of idiosyncrasies, $\sigma$.

We have represented in Figs.~\ref{logd04} and \ref{logd03} the time
evolution of the number of adopters exhibiting the oscillations, when
they happen, or the convergence to a fixed point when there are no
oscillations. When performing Monte Carlo simulations there are no
oscillations in none of the cases. Figure \ref{logres} summarizes the
numerical results for the logistic distribution. Red curves (dashed)
correspond to the final number of adopters (fixed points) when
performing MC simulations, while black curves correspond to Parallel
Dynamics (PD). In the later case, oscillations may be observed above a
critical value of the fraction of contrarians. When there are no
oscillations, the results for PD and MC simulations coincide. When
oscillations are present, PD results are different from MC results,
and the plot shows bot the extreme amplitude of the oscillations and,
in between, the average value of the number of adopters.

\subsection{Analytic results}

The dynamics of adoption, given by equation (\ref{eq:dynamics}), is
\begin{equation}
\label{eq:dynamicsLogistic}
n(t+1) = (1-f) {\cal F}(d+ n(t)) + f {\cal F}(d- n(t))
\end{equation}
where ${\cal F}(u)$ is the cumulative distribution
\begin{equation}
{\cal F}(u) = \int_{-\infty}^u {\cal P}(x) dx = \frac{1}{1+e^{-2\beta
    u}}.
\end{equation}

\begin{figure*}[htbp]
	\centering \includegraphics[width=1.0\textwidth]{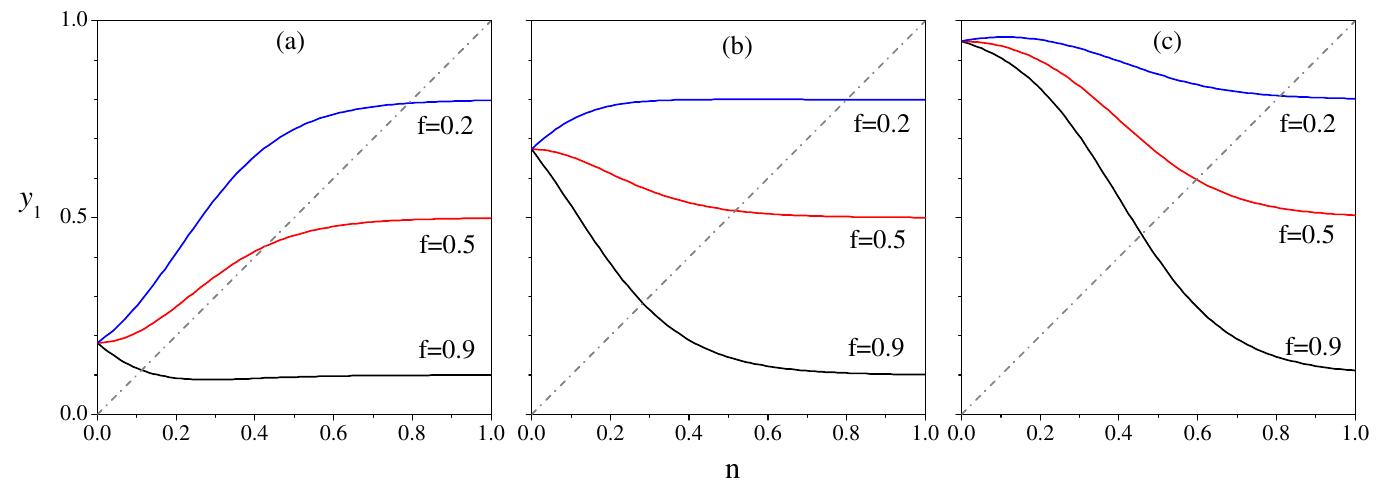}
	\caption{Fixed points of $y_1(n)$. The fixed points correspond
          to the intersections of $y_1(n)$ and $y_2(n)$ (indicated by
          the dot-dashed gray line in the figure). When the absolute
          value of the derivative is lower than one, the solutions are
          stable and correspond to a fixed point of the dynamics.  The
          three panels show different cases for different values of
          the parameter $d$ of the logistic distribution of
          idiosyncrasies: (a) For $d=-0.2$ the derivatives at the
          intersections are always $ |y_1'| <1$, thus no oscillations
          are expected.  (b) When $d=0.1$ three possible stable
          intersections appear in each case, and the values roughly
          correspond to the numerical results plotted on
          Fig.~\ref{logres}(a).  (c) When $d=0.4$ and $f=0.2$ the
          stable solution correspond to $n \approx 0.8$ that coincides
          with the numerical solution (See Fig.~\ref{logres}(b)).  For
          $f=0.5$, $n \approx 0.5$ that also coincides with both PD
          and MC simulations. Finally, for $f=0.9$ the solution is
          unstable ($ |y_1'| >1$). However the fixed point corresponds
          to the value obtained with MC simulations (see
          Fig.~\ref{logres}), while PD simulations indicated the
          existence of oscillations. }\qquad\qquad
	\label{logana01}
\end{figure*}

The fixed points may be obtained by solving the transcendental
equation

\begin{equation}
n = y_1(n)=\frac{1-f}{1+e^{-2\beta (d+n)}} + \frac{f}{1+e^{-2\beta
    (d-n)}}
\end{equation}
through the intersections of the function $y_1(n)$ with the line
$y_2(n)=n$. Fig. \ref{logana01} presents some examples for different
parameter values.

\begin{figure}[htbp]
		\centering
                \includegraphics[width=0.4\textwidth]{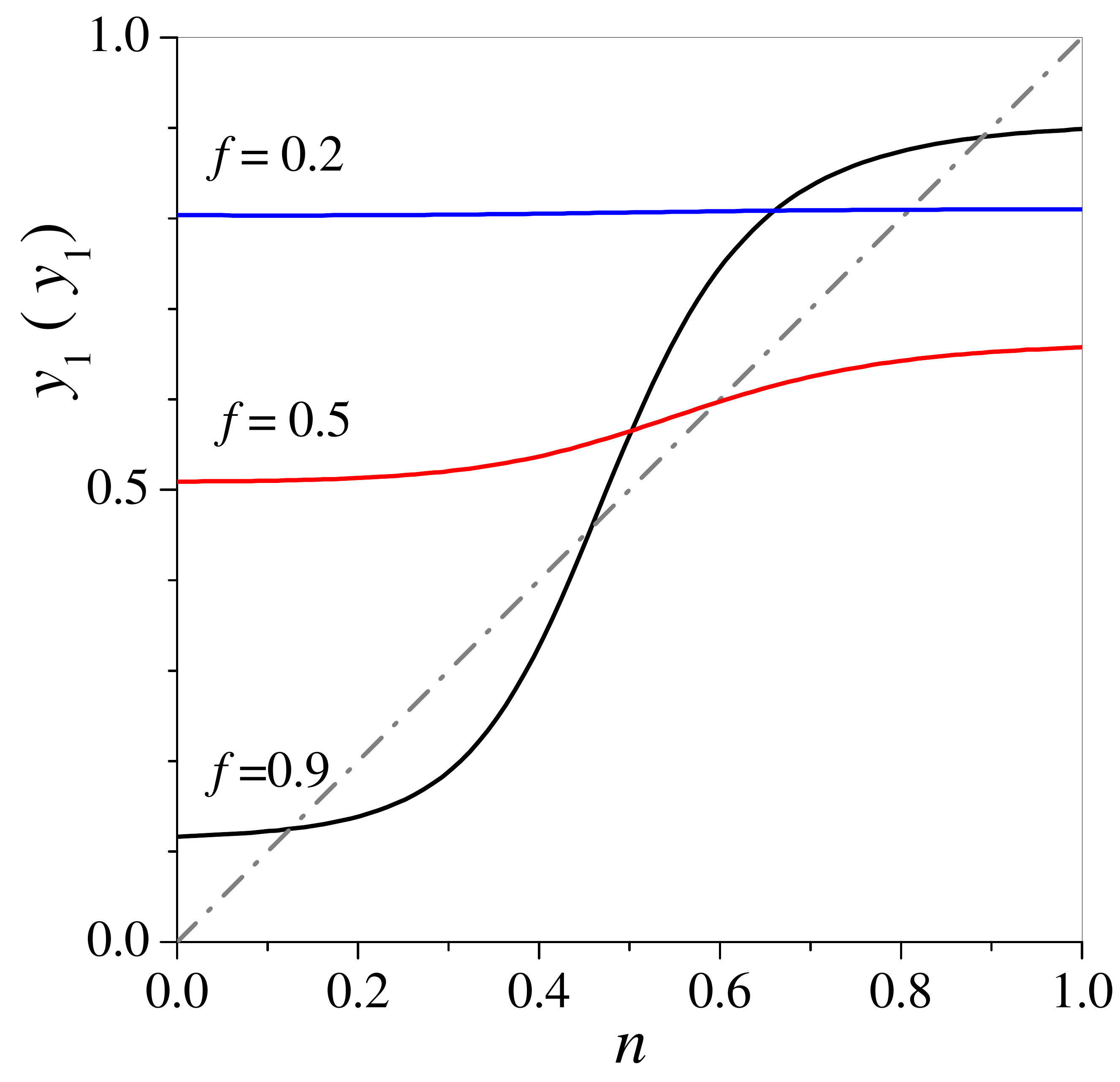}
		\caption{Fixed points of $y_1(y_1(n))$ with $y_2(n)$
                  (gray dot-dashed line). We have plot just the case
                  with $d=0.4$. It is possible to observe that for
                  $f=0.2$ and $f=0.5$ there is just one intersection,
                  that corresponds to the stable solutions previously
                  obtained. For $f=0.9$ there are three
                  intersections. The middle one corresponds to the
                  fixed point of $y_1(n)$ while the other two
                  represent the extremes of the oscillations. These
                  extreme values are approximately $0.12$ and $0.9$
                  and correspond to the extreme value of the
                  oscillations in the PD simulations, see
                  Fig. \ref{logd04}(c).} \qquad \qquad
			\label{logana02}
\end{figure}

Fig. \ref{logana01} represents a plot of $y_1(n)$ and the line
$y_2(n)=n$. The intersections correspond to the fixed points and are
stable solutions provided that $ y_1' \equiv \frac{d y_1}{d n} < 1$
. However, solutions with $| y_1'| \equiv |\frac{d y_1}{d n}| >1$ are
unstable, and we are then obliged to consider a second iteration,
i.e., $y_1(y_1(n))$. The solutions for this second iteration are
represented on Fig. \ref{logana02}: if more than one intersection is
present, the upper and lower intersections correspond to the extreme
value of the oscillations.

The comparison between numerical and analytical solutions is discussed
in detail in the caption of Figs. \ref{logana01} and
\ref{logana02}. We find a very good agreement of both solutions, then,
there is no need of further discussion of this point. We will
concentrate in the next section in the discussion of the results and
comparison with a previous model~\cite{GoncalvesLagunaIglesias12}.


\section{Discussion and Conclusions}
\label{conclu}

\begin{figure}[htbp]
	\centering
        \includegraphics[width=0.45\textwidth]{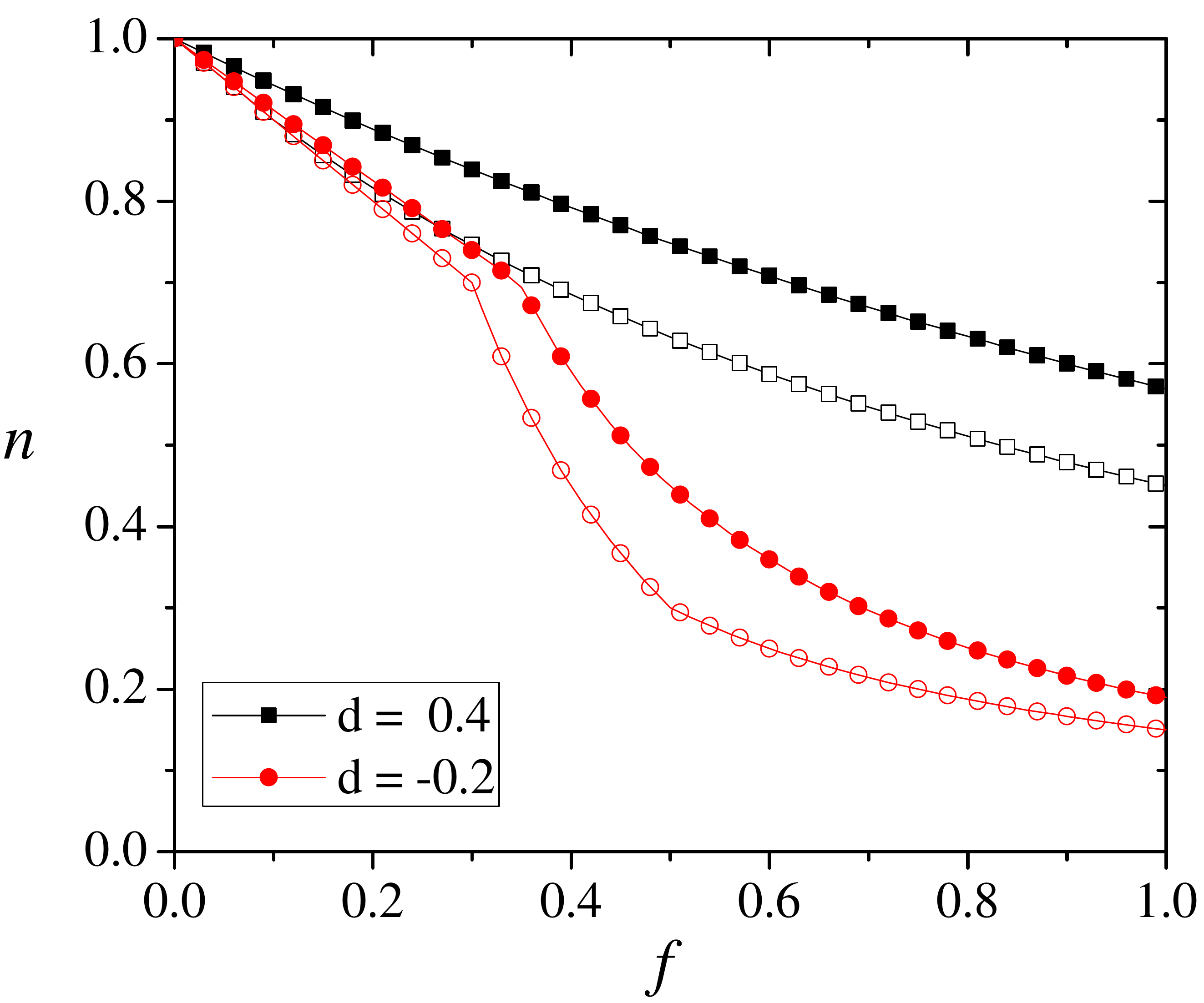}
	\caption{Comparison between the results of
          ref.~\cite{GoncalvesLagunaIglesias12} (without repentants)
          and the present ones with repentants: final number of
          adopters for two values of $d$ ($d=0.4$, squares, and
          $d=-0.2$, circles). Filled symbols correspond to no
          repentants and open ones with repentants (present
          contribution).  Pairs of curves display similar behavior
          with always lower values of adoption for the case with
          repentants.}  \qquad \qquad
	\label{compare}
\end{figure}

One of the interesting points of the present contribution is that the
temporal behavior of the adoption of innovations is very sensitive to
the width of the distribution of the resistance to adopt $u_i$.  In
the case of a wide uniform distribution, as the one utilized in
ref~\cite{GoncalvesLagunaIglesias12}, oscillations appear as a
transient state but in the end the system converges to a fixed point.
On the other side, with narrower uniform distributions of $u_i$,
sustained oscillations appear, which are produced by the contrarians,
whereas mimetic agents hardly change their decision.  It could be
argued that this kind of distribution is not representative of social
systems. However, when the distribution is bell shaped, as it is the
case of the logistic distribution presented in section \ref{logistic},
we obtain similar results. That is, stable long term oscillations may
appear for intermediate values of the advertising, $d $, or a large
fraction of contrarians, $f$, as it is evident in Fig.~\ref{umbral}.
While a high number of contrarians may be unreal considering a novel
technology, that could be the case regarding brand choices, for
example ({\em iPhone} vs. {\em Samsung}).  In any case,
Fig.~\ref{umbral} shows that oscillations may appear with a relatively
low fraction of contrarians, provided the advertising is strong: see
for example that for a narrow distribution of idiosyncrasies
($\sigma=0.05$) and for $d\approx 0.95$ the threshold is of the order
of $f \approx 0.15$.  Also, the coexistence of contrarians with the
possibility of changing decisions makes the final total number of
adopters lower than in the case with no
regrets~\cite{GoncalvesLagunaIglesias12}.  To check this point we have
represented in Fig.~\ref{compare} the present results for the uniform
distribution of $u_i$ together with those of
Ref.~\cite{GoncalvesLagunaIglesias12}. It is possible to see that the
shape is similar in both cases, but when the decision is
``reversible'' the final adoption is lower than when not. To produce
this comparison we considered the uniform distribution of
idiosyncrasies because it was the one used in
Ref.~\cite{GoncalvesLagunaIglesias12}.

The presence of repentants and contrarians have the effect of reducing
the final number of adopters. Also, oscillations may appear when the
distribution of resistances to adopt is smaller than $1$, {\em i.e.}
smaller than the social interaction, $J$.  Also, such cycles are only
possible if both, contrarians and repentants, are present.  Concerning
the oscillations, they are interesting but somehow artificial, as they
mainly arise from contrarians that regret their decision. In any case
one expect that the first oscillations are the important ones, as it
is not plausible that agents continue to change their minds all the
time.  Regarding the fact that the inclusion of contrarians and
repentants reduces significantly the final fraction of adopters, a
direct consequence is that a stronger advertising campaign will be
needed if it seeks to impose innovation. Work in progress include the
presence of impulsive agents, i.e., people that can change their
technology without assessing whether it is convenient. Preliminary
results show that the introduction of such agents in the model
accelerates the adoption process and increases the total number of
adopters. In other words, a small number of impulsive agents in a
society could be as efficient as a strong advertising.

We are also considering a dynamic distribution of idiosyncrasies, the
effect of distributing the agents on a network, and a non-linear term
of social interaction that may describe the effects of fashion: people
adopt a new fashion when there are a few followers but abandon when
the number of adopters increases.

Concluding, we would like to point out that, despite its simplicity,
this model reproduce some common features of the innovation adoption
process while allowing the study of more realistic cases.

\section{Acknowledgments}
\label{ackn}
One of us (JRI) acknowledges financial support of Brazilian agency
CNPq and Argentinian agency CONICET. He also thanks the kind
hospitality of the IFIMAR (Instituto de F\'{\i}sica de Mar del Plata)
during 2015-2016. MFL is member of CONICET.

\end{document}